\documentclass[12pt]{article}
\usepackage{graphicx}
\DeclareGraphicsExtensions{.ps,.eps}
\usepackage{amsmath}
\usepackage{amsthm}
\theoremstyle{definition}
\newtheorem{thm}{Theorem}
\newtheorem{cor}[thm]{Corollary}
\makeatletter
\renewcommand\@biblabel[1]{#1.}
\makeatother
\begin{document}
%\begin{titlepage}
\title{A new Information Metric and a possible higher bound for a class of measurements in the Quantum Estimation Problem.}
\author{Demetris P.K. Ghikas,\thanks{ghikas@physics.upatras.gr, Tel: +302610-997460, FAX: +302610-997617}
\and Fotios D. Oikonomou \thanks{One of the authors (F.O.) wishes to thank the Greek State Scholarships Foundation for a research scholarship.}
\and Department of Physics
University of Patras,
Patras 26500, Greece}
\maketitle
\begin{abstract}
Information metrics give lower bounds for the estimation of parameters. The Cencov-Morozova-Petz Theorem classifies the monotone quantum Fisher metrics. The optimum bound for the quantum estimation problem is offered by the metric which is obtained from the symmetric logarithmic derivative. To get a better bound, it means to go outside this family of metrics, and thus inevitably, to relax some general conditions. In the paper we defined logarithmic derivatives through a phase-space correspondence. This introduces a function which quantifies the deviation from the symmetric derivative. Using this function we have proved that there exist POVMs for which the new metric gives a higher bound from that of the symmetric derivative. The analysis was performed for the one qubit case. 
\end{abstract}
%\pacs{03.67-a, 03.65-Aa, 02.50 Sk, 02.40 Hm}
%\keywords{information metrics, quantum estimation problem,\\ estimation inequalities, phase-space correspondence}
%\submitto{\JPA}
%\maketitle
\textbf{Mathematics Subject Classification.} 81Q99, 53Z05, 62F12\\
\textbf{Keywords} information metrics, quantum estimation problem, quantum measurments, estimation bounds.

\section{Introduction}

\indent
In the classical parameter estimation theory a fundamental role is played by the Cramer-Rao inequality. According to the latter the variance-covariance of an unbiased estimator is bounded from below by the inverse of the Classical Fisher Information Matrix, which is constructed from the logarithmic derivatives of the parametric statistical model. The Fisher Information Matrix provides a metric on the information manifold of the probability distributions of the model. A basic property of such a metric on a statistical manifold is its monotonicity under stochastic (information loosing) transformations. This means that  the distributions are closer w.r.t. such metrics or become less distinguishable under such transformations. The well known theorem of Cencov \cite{cencov} states that the Classical Fisher Metric is the unique metric with the above monotonicity property. In the Quantum Parameter Estimation Problem there are new aspects in the parameter estimation inequalities. First a parameter of a quantum state is estimated using appropriate quantum measurements which provide classical probability distributions. These distributions give the classical Fisher metrics with   corresponding Cramer-Rao inequalities. Second the question of the existence of metrics which are independent of the measurements and characteristic only of the quantum states is related to the existence of Quantum Fisher Metrics with the monotonicity property under completely positive maps of the states. The Theorem of Cencov-Morozova-Petz \cite{petz1,petz2} gives the general form of all monotone metrics. With respect to these metrics there is a hierarchy of Cramer-Rao type inequalities which provide bounds for the variance-covariance of unbiased estimators. While the highest bound is given by the Classical Fisher metric of the measurement probability, the quantum monotone metrics give lower bounds \cite{hayashi1}. The highest among them is given by the symmetric logarithmic derivative (SLD) and the lowest by the right logarithmic derivative (RLD). There is an extensive literature on the saturation of these inequalities for finite ensembles and in the asymptotic regime \cite{nielsen1,gill1,wiseman1}. We posed a question in the opposite direction. While the saturation of the inequalities cannot be achieved for all measurements and all types of quantum states, we asked whether there exist higher bounds for a class of measurements that give a less informative estimation. This would be useful in classifying the effectiveness of certain measurements which may actually be realized. Existence of higher bounds means less accuracy in parameter estimation for certain measurements. Given the fact that the SLD gives the highest bound which, anyway, is smaller than the classical Fisher bound, a higher bound, in the spirit of quantum metrics, can be derived if one goes outside the family of metrics resulting from the Cencov-Morozova-Petz Theorem. Since the non-uniqueness of the quantum information metrics is related to the non-commutativity of the logarithmic derivatives, it seemed natural to construct logarithmic derivatives indirectly through a phase-space correspondence. That is, for the given quantum state, we took  the logarithmic derivatives of the corresponding Husimi function and mapped them back as operators in the original Hilbert space. Using these operator valued objects we derived a metric. But, as expected, at least for the case of one qubit, the form is not that of the Petz family of metrics. This means that our metric is not manifestly monotone. Nevertheless we investigated what could be its role in the hierarchy of Cramer-Rao type inequalities. We introduced a function which in some sense quantifies the deviation of the new metric from the SLD metric. Using this function we proved that there exist measurements for which the mean square error is bounded from below, with a bound even higher than  the classical Fisher bound. This result could be useful in the investigation of the effectiveness of certain estimation procedures. \\ 
In the paper we first give a short introduction on the classical and quantum estimation problems with the corresponding inequalities. Then we present the necessary tools for the phase-space correspondence. Next we construct the new logarithmic derivatives and the new quantum metric. Using this we state a theorem concerning the new bound and apply it to an example of one qubit. The proof is given in the appendix. Finally we discuss open problems and further possibilities of this approach.        
\section{The classical and quantum parameter estimation problem}
\indent
Here we present only the necessary concepts in order to establish the notation. We refer to the bibliography for the details \cite{amari1,hayashi1}. Let
\begin{equation}
S=\{p_{\xi}=p(x;\xi) | \xi = [\xi^{1},...,\xi^{n}]\in \Xi\}
\end{equation}
be a parametric family of probability distributions on $\mathcal{X}$. This is an n-dimensional parametric statistical model \cite{amari1}. Given the N observations $x_{1},...,x_{N}$, the Classical Estimation Problem concerns the statistical methods that may be used to detect the true distribution, that is to estimate the parameters $\xi$. To this purpose, an appropriate estimator is used for each parameter. These estimators are maps from the parameter space to the space of the random variables of the model. The quality of the estimation is measured by the variance -covariance matrix $V_{\xi}=[v_{\xi}^{ij}]$ where
\begin{equation}
v_{\xi}^{ij}= E_{\xi}[(\hat{\xi^{i}}(X)-\xi^{i})(\hat{\xi^{j}}(X)-\xi^{j})]
\end{equation} 
Suppose that the estimators are unbiased, namely
\begin{equation}
E_{\xi}[\hat{\xi}(X)]=\xi \quad  \quad \forall \xi \in \Xi
\end{equation}
Then a  lower bound for the estimation error is given by the Cramer-Rao inequality
\begin{equation}
V_{\xi}(\hat{\xi})\geq G(\xi)^{-1}
\end{equation}
where $G(\xi)=[g_{ij}(\xi)]$ 
\begin{equation}
g_{ij}(\xi)=E_{\xi}[\partial_{i}l(x;\xi)\partial_{j}l(x;\xi)]
\end{equation}
the Classical Fisher Matrix with 
\begin{equation}
l_{\xi}=l(x;\xi)=logp(x;\xi)
\end{equation}
the score function.
As it has been shown the Fisher Matrix provides a metric on the manifold of classical probability distributions. This metric, according to the theorem of Chencov \cite{cencov}, is the unique metric which is monotone under the transformations of the statistical model. This means that if the map $F : \mathcal{X} \to \mathcal{Y}$ induces a model $S_{F}=\{q(y;\xi)\}$ on $\mathcal{Y}$ then 
\begin{equation}
G_{F}(\xi)\leq G(\xi)
\end{equation} 
That is, the distance of the transformed distributions is smaller than that of the original distributions. Thus monotonicity of the metric is intuitively related to the fact that in general we loose distinguishability of the distributions after applying any transformation to the information.
\indent \\
The case of Quantum Estimation Problem differs in two fundamental aspects from the classical one. First the parameters to be estimated are classical parameters of the apparatuses that prepare the quantum system to a given state. These parameters are measured through an interaction of the system with appropriate measuring devices. The measurements give classical distributions which must be estimated as in the classical case, and thus the estimation errors obey the corresponding Cramer-Rao inequalities. The efficiency of this procedure depends on the effectiveness of the measurements used. The immediate question is what is the ultimate bound for the estimation error. This bound must depend only on the state of the system and as it turned out it is given by the inverse of a quantity termed Quantum Fisher Information Metric \cite{amari1,hayashi1}. Let $\Pi =\{\pi(x)\}$ be the measurements used to estimate the state $\rho_{\xi}$ of the system. Then the probability distribution of the measurements is $p(x|\xi)=tr[\rho_{\xi}\pi(x)]$. The corresponding classical Fisher information metric  $G_{\Pi}(\xi)$ satisfies the inequality
\begin{equation}
G_{\Pi}(\xi)\leq G_{q,s}(\xi)
\end{equation} 
where $G_{q,s}(\xi)$ is the Quantum Fisher Information Metric of \cite{wiseman1,hayashi1}, while the index \textit{s} means \textit{symmetric} and indicates the way this metric is constructed, that is the way the logarithmic derivative is generalized to an appropriate operator. More specifically this operator valued quantity is defined implicitly by the equation

\begin{equation}
\frac{\partial \rho_{\xi}}{\partial \xi^{i}}=\frac{1}{2}(\rho_{\xi}L_{i}+L_{i}\rho_{\xi})
\end{equation} 

and then $G_{q,s}(\xi)$ is defined by the equation

\begin{equation}
g_{i,j,q,s}(\xi)=tr(\rho_\xi \frac{1}{2}(L_{i}L_{j}+L_{j}L_{i}))
\end{equation}

From now on we write $g_{q,s}(\xi)$ as $g_{SLD}(\xi)$ denoting that it is constructed through a \textit{symmetric logarithmic derivative}.
The SLD metric is monotone under completely positive maps, but it is not unique. This fact is the second fundamental difference from the classical case. There are various approaches for the definition and construction of Quantum Fisher Information Metrics \cite{amari1,hayashi1,petz1,petz2}. The Cencov-Morozova-Petz Theorem \cite{petz1,petz2} gives a general characterization of monotone metrics using operator monotone functions \cite{petz1,petz2,kubo1}. For these metrics we have the inequalities

\begin{equation}
  V_{\xi}(\hat{\xi})\geq G_{\Pi}(\xi)^{-1} \geq G_{SLD}(\xi)^{-1} \geq G_{M}(\xi)^{-1} \geq G_{RLD}(\xi)^{-1}
\end{equation}   
where $G_{M}(\xi)$ is any monotone metric and $G_{RLD}(\xi)$ is the  metric corresponding to the {\textit right logarithmic derivative} defined as
\begin{equation}
\frac{\partial \rho_{\xi}}{\partial \xi^{i}}=\rho_{\xi}L_{i}
\end{equation}
For the case of one qubit that we analyze here, the monotone metrics have the general form \cite{petz1,petz2}
\begin{equation}
ds^{2}=\frac{1}{1-r^{2}}dr^{2}+\frac{1}{1+r}g(\frac{1-r}{1+r})dn^{2}
\end{equation}
where $g(t)=\frac{1}{f(t)}$ and $f(t)$ satisfies $f(t)=tf(t^{-1})$.
The above inequalities satisfied by the quantum information metrics lead to the conclusion that, as far as the estimation problem is concerned, only the SLD metric has any practical use, though there are some cases where other metrics play a role \cite{hayashi1}. A great part of the literature on the estimation problem deals with the question of saturation of these inequalities mainly in the asymptotic case \cite{hayashi1,nielsen1,gill1}.
In the present work we pose a question in the opposite direction. We ask whether there exit measurements for which there is a higher bound in the estimation error. In the sequel, after introducing our phase-space construction of the logarithmic derivatives, we prove the existence of such measurements with a higher lower bound even higher than that of the classical Fisher metric of the corresponding measurement.
\section{Phase-space correspondence}
\indent \\
Since our interest is on the question of the existence of measurements with higher lower bounds of estimation errors, we restrict our analysis to the one qubit case. Then the Hilbert space is 2-dimensional and the information manifold 3d \cite{bengtsson1}. The state of one qubit is given by the density operator
\begin{equation}
\rho = \frac{1}{2}(\boldsymbol{1}+r\hat{n}\cdot \boldsymbol{{\vec\sigma}})
\end{equation}
where $\boldsymbol{\vec{\sigma}}=(\boldsymbol{\sigma_{1}},\boldsymbol{\sigma_{2}},\boldsymbol{\sigma_{3}})$ and $\hat{n}=(sin\theta cos\phi,sin\theta sin\phi,cos\theta)$ with $0 \leq r \leq 1$ , $0 \leq \theta \leq \pi$ , $0 \leq \phi \leq 2\pi$. Explicitly we have
\begin{equation}
\rho(r,\theta,\phi)= \frac{1}{2}
\left(
\begin{array}{ccc}
%\eqalign
1+rcos\theta         &  re^{-i\phi}sin\theta  \\
re^{i\phi}sin\theta  &  1-rcos\theta
\end{array}
\right)
\end{equation}
$r,\theta,\phi$ are the coordinates of the information manifold for the qubit. The dynamical symmetry group for the system is $SU(2)$ and the phase space is the unit sphere $S^{2}=SU(2)/U(1)$. Let $\Omega (\theta,\phi)$ be a point on $S^{2}$. The Weyl Rule
\begin{equation}
F_{A}(\Omega;s)=tr[A\Delta(\Omega;s)]
\end{equation}
is a map of an operator A acting on the Hilbert space to a function on the phase space. $\{\Delta(\Omega;s)\}$ is a parametric family of operators defined on the phase space, and called Stratonovich-Weyl Kernels \cite{brif1}.
The inverse map is
\begin{equation}
A=\int\limits_{S^{2}}d\mu(\Omega)F_{A}(\Omega;s)\Delta(\Omega;-s)
\end{equation}
where here $ d\mu(\Omega)=sin\theta d\theta d\phi/2\pi $.
We have
\begin{equation}
\Delta(\theta,\phi;s)=\sqrt{2\pi}\sum_{l=0}^{1}<1/2,1/2;l,0|1/2,1/2>^{-s}\sum_{m=-l}^{l}D_{lm}Y_{lm}^{*}(\theta,\phi)
\end{equation}
where
\begin{equation}
D_{lm}=\sqrt{\frac{2l+1}{2}}\sum_{k,q=-1/2}^{1/2}<1/2,k;l,m|1/2,q > |q><k|
\end{equation}
or explicitly
\begin{equation}
\Delta(\theta,\phi,s)=\frac{1}{2}
\left(
\begin{array}{ccc}
1+3^\frac{1+s}{2}cos\theta         &  3^\frac{1+s}{2}e^{-i\phi}sin\theta  \\
3^\frac{1+s}{2}e^{i\phi}sin\theta  &  1-3^\frac{1+s}{2}cos\theta
\end{array}
\right)
\end{equation}
For $A=\rho$ we get the quasiprobability distributions (QPD) $P(\Omega;s)=F_{\rho}(\Omega;s)$. For $s=-1$ we get the Husimi Q-function, which is non-negative and thus a probability distribution.
We have
\begin{equation}
 Q(r,\theta,\phi;\theta_{1},\phi_{1})
=tr[\rho(r,\theta,\phi)\Delta(\theta_{1},\phi_{1},-1)]\\=\frac{1}{2}(1+rcos\theta_{1}cos\theta+rcos(\phi_{1}-\phi)sin\theta_{1}sin\theta)
\end{equation}
\indent \\
Now having a genuine probability distribution associated directly to the state of the system, and not specific to a measurement, we may construct a Fisher metric.
\begin{equation}
g_{ij,Husimi}=E_{Q}[\partial_{i}logQ(\xi;\Omega)\partial_{j}logQ(\xi;\Omega)]
\end{equation}

 This was proposed as a way to avoid the non-commutativity difficulties of the direct Hilbert space constructions \cite{ghikas1,ghikas2}. In \cite{ghikas2} a classical Fisher metric was constructed for a squeezed state. As it turned out, this metric was proportional to the SLD one. In \cite{slater1} using this phase space approach a metric was given in the form
\begin{equation}
ds_{Husimi}^{2}=\frac{-2r-log\left(\frac{1-r}{1+r}\right)}{2r^{3}}dr^{2}+\frac{2r+(1-r^{2})log\left(\frac{1-r}{1+r}\right)}{4r^{3}}dn^{2}
\end{equation}
This metric is manifestly non-monotone, as it differs from the standard form of the monotone metrics \cite{petz1,petz2}.
The conclusion is that, though the construction of the classical Fisher metric through the phase-space distribution is very natural, it turned out that either it gives no new metric, or it gives one manifestly outside the monotone family. Motivated by these results we looked at an alternative route, that of a construction of quantum logarithmic derivatives through the phace-space correspondence. 
\section{A new quantum information metric}
\indent\\
We propose the following equally natural construction: Starting from the quantum state, construct the phase-space distribution (Husimi function) and take partial derivatives with respect to the relevant parameters of the model. These are functions on the phase-space. Then using the Stratonovich-Weyl Kernels map these functions back to Hilbert space operators. Finally use these operators to construct, in the standard way, the Quantum Fisher Metric.
We present our results for the one qubit case.\\
We have for the new operator logarithmic derivative
\begin{equation}
L_{i}=L_{i}(r,\theta,\phi)=\int_{\theta_{1}=0}^{\pi}\int_{\phi_{1}=0}^{2\pi}\partial_{i}logQ(r,\theta,\phi,\theta_{1},\phi_{1})\Delta(\theta_{1},\phi_{1},1)\frac{sin\theta_{1}d\phi_{1}d\theta_{1}}{2\pi}
\end{equation}
We get
\begin{equation}
L_{1}=\frac{1}{2r^3}\left(2r+log\left(\frac{1-r}{1+r}\right)\right)
\left(
\begin{array}{ccc}
r-3cos\theta         &  -3e^{-i\phi}sin\theta  \\
-3e^{i\phi}sin\theta   &  r+3cos\theta
\end{array}
\right)
\end{equation}

\begin{equation}
L_{2}=\frac{3}{4r^2}\left(2r+(1-r^{2})log\left(\frac{1-r}{1+r}\right)\right)\left(
\begin{array}{ccc}
-sin\theta         &  e^{-i\phi}cos\theta  \\
e^{i\phi}cos\theta   &  sin\theta
\end{array}
\right)
\end{equation}

\begin{equation}
L_{3}=\frac{3}{4r^2}\left(2r+(1-r^{2})log\left(\frac{1-r}{1+r}\right)\right)\left(
\begin{array}{ccc}
0        &  -ie^{-i\phi}sin\theta  \\
ie^{i\phi}sin\theta   &  0
\end{array}
\right)
\end{equation}
From these we have
\begin{equation}
ds^{2}=\frac{9-5r^{2}}{4r^{6}}\left(-2r-log\left(\frac{1-r}{1+r}\right)\right)^{2}dr^{2}+\frac{9}{16r^{6}}\left(2r+(1-r^{2})log\left(\frac{1-r}{1+r}\right)\right)^{2}dn^{2}
\end{equation}
This is our new quantum Fisher metric, which is manifestly non-monotone. Using this metric we construct a new bound for the estimation error which for some measurements is even higher than that of the corresponding classical Fisher bound.

\section{Existence of a new bound}
\indent \\
We introduce three operators which quantify the deviation of our new metric from the SLD metric
\begin{equation}
h_{i}=\frac{\partial\rho}{\partial\xi^{i}}-\frac{1}{2}\left(\rho L_{i}+L_{i}\rho\right) 
\end{equation}

\begin{equation}
h_{1}=\frac{4r^{2}+2rlog\left(\frac{1-r}{1+r}\right)}{4r^{3}}
\left(
\begin{array}{ccc}
1      &  0  \\
0      &  1
\end{array}
\right)
+\frac{6r+(3-r^{2})log\left(\frac{1-r}{1+r}\right)}{4r^{3}}
\left(
\begin{array}{ccc}
cos\theta                  &    sin\theta e^{-i\phi}  \\
sin\theta e^{i\phi}       &    -cos\theta
\end{array}
\right)
\end{equation}
\begin{equation}
h_{2}=
\frac{6r-4r^{3}+3(1-r^{2})log\left(\frac{1-r}{1+r}\right)}{8r^{2}}
\left(
\begin{array}{ccc}
sin\theta                  &    -cos\theta e^{-i\phi}  \\
-cos\theta e^{i\phi}       &    -sin\theta
\end{array}
\right)
\end{equation}
\begin{equation}
h_{3}=
\frac{6r-4r^{3}+3(1-r^{2})log\left(\frac{1-r}{1+r}\right)}{8r^{2}}
\left(
\begin{array}{ccc}
0                          &    isin\theta e^{-i\phi}  \\
-isin\theta e^{i\phi}       &    0
\end{array}
\right)
\end{equation}
These operators are involved in the construction of the bound. Here we define the main quantities and present the main results. The detailed proof of the theorem is presented  in the appendix.
\subsection{The General Bound}
We suppose that we approximately know the value of the parameter $\xi$, with an error $\epsilon$ and that, to be general, the POVM measurement depends on $\xi^{'}$ i.e. $\Pi=\Pi(\hat{\xi},\xi{'})$, where $\xi{'}=\xi+\epsilon$. Then the classical probability distribution to be estimated is $q_{\epsilon}(\hat{\xi}|\xi)=tr[\rho(\xi)\Pi(\hat{\xi},\xi{'})]$  
We assume that the estimation is unbiased, namely
\begin{equation}
\xi^{i}=E_{q_{\epsilon}}(\hat{\xi^{i}}|\xi)\equiv \int \hat{\xi^{i}}q_{\epsilon}(\hat{\xi}|\xi)d\hat{\xi}
\end{equation}
Let $Y^{t}=(y_{1},y_{2},y_{3})$ and $Z^{t}=(z_{1},z_{2},z_{3})$. We define the quantities
\begin{equation}
T_{\xi} = \boldsymbol{1}\sum_{j=1}^{3}y_{j}(\hat{\xi^{j}}-\xi^{j}) \quad T_{h}=T_{h}(\xi)=\sum_{j=1}^{3}z_{j}h_{j}(\xi) 
\end{equation}
\begin{equation}
A_{k}^{j}(\xi)=\xi^{j}\int tr\left( \rho(\xi)\frac{\partial\Pi(\hat{\xi},\xi+\epsilon)}{\partial\xi^{k}}\right)d\hat{\xi}
\end{equation}

\begin{equation}
B_{k}^{j}(\xi)=\delta_{k}^{j}-\int \hat{\xi^{j}}tr\left( \rho(\xi)\frac{\partial\Pi(\hat{\xi},\xi+\epsilon)}{\partial\xi^{k}}\right)d\hat{\xi}
\end{equation}
We have
\begin{thm}
Let $V_{q_{\epsilon}}=\{v_{q_{\epsilon}}^{ij}\}$ 
\begin{eqnarray}
 v_{q_{\epsilon}}^{ij}&=&E_{q_{\epsilon}}\left[\left(\hat{\xi^{i}}-\xi^{i}\right)\left(\hat{\xi^{j}}-\xi^{j}\right)  |\xi\right] \\
&\equiv& \int \left[\left(\hat{\xi^{i}}-\xi^{i}\right)\left(\hat{\xi^{j}}-\xi^{j}\right)q_{\epsilon}(\hat{\xi}|\xi)\right]d\hat{\xi}
\end{eqnarray}
the covariance matrix w.r.t. $q_{\epsilon}(\hat{\xi}|\xi)$. \\
Then we have
\begin{equation}
\left( Y^{t}V_{q_{\epsilon}}Y \right)\left(Z^{t}G Z \right)\geq \left|Y^{t}\left(A+B\right )Z - tr\int T_{\xi}T_{h}\Pi(\hat{\xi},\xi + \epsilon)d\hat{\xi}   \right |^{2}
\end{equation}
\end{thm}
The proof can be found in the Appendix. It is based on a similar Theorem of \cite{helstrom1}.\\
The above inequality is general, and it holds for an arbitrary measurement. The crucial point is that it depends on the operators h which depend only on the state, and not on the measurement. This means that for a given state we may extremize w.r.t. the measurements in order to uncover a possible new lower bound for a class of measurements. Since the general case is quite involved, and since our purpose is to show that there exist such bounds, we analyze in the next section the case of one unknown parameter. 
\subsection{One Dimensional case}
For the one qubit case the parameters are $\xi = (r,\theta,\phi)$. Considering $\phi$ to be the single unknown parameter the theorem states that
\begin{equation}
v_{q_{\epsilon}}g \geq \left |a(\phi)+b(\phi)-tr \int_{\phi+\epsilon - \pi}^{\phi + \epsilon + \pi}\left(\hat{\phi}-\phi\right)h_{3}\Pi(\hat{\phi},\phi+\epsilon)d\hat{\phi}\right |^{2}
\end{equation}
where
\begin{equation}
g=\frac{9}{16r^{4}}\left(2r+(1-r^{2})log\frac{1-r}{1+r}\right)^{2}sin^{2}\theta
\end{equation}
\begin{eqnarray}
a(\phi)&=&\phi \int_{\phi+\epsilon - \pi}^{\phi + \epsilon + \pi} tr \left(\rho(\phi)\frac{\partial \Pi(\hat{\phi},\phi+\epsilon)}{\partial \phi}\right)d\hat{\phi} \\
b(\phi)&=&1-\int_{\phi+\epsilon - \pi}^{\phi + \epsilon + \pi}\hat{\phi}  tr \left(\rho(\phi)\frac{\partial \Pi(\hat{\phi},\phi+\epsilon)}{\partial \phi}\right)d\hat{\phi}
\end{eqnarray}
A general POVM in the present case has the form
\begin{equation}
\Pi(\hat{\phi},\phi)=
\left(
\begin{array}{ccc}
x^{11}(\hat{\phi},\phi)    &     x^{12}(\hat{\phi},\phi)  +  iy^{12}(\hat{\phi},\phi)   \\
x^{12}(\hat{\phi},\phi)  -  iy^{12}(\hat{\phi},\phi)        &    x^{22}(\hat{\phi},\phi)  
\end{array}
\right)
\end{equation}
Then we get
\begin{equation}
v_{q_{\epsilon}}g \geq \left (a(\phi)+b(\phi)-2h_{3r}(r)sin\theta  \int_{\phi+\epsilon - \pi}^{\phi + \epsilon + \pi}\left(\hat{\phi}-\phi\right)(sin\phi x^{12}+cos\phi y^{12})d\hat{\phi}\right )^{2}
\end{equation}
where
\begin{equation}
h_{3r}=
\frac{6r-4r^{3}+3(1-r^{2})log\left(\frac{1-r}{1+r}\right)}{8r^{2}}
\end{equation}
We note that the quantity inside the absolute value, according to the proof of the Theorem 1, is real.
Now for measurements independent of the estimated parameter we can state the result as a Corollary
\begin{cor}
Let the measurement be independent of the estimated parameter $\phi$, then since $a(\phi)=0$ and $b(\phi)=1$ the inequality simplifies to 
\begin{equation}
v_{q_{\epsilon}}g \geq \left( 1 -2h_{3r}(r)sin\theta  \int_{\phi+\epsilon - \pi}^{\phi + \epsilon + \pi}\left(\hat{\phi}-\phi\right)(sin\phi x^{12}+cos\phi y^{12})d\hat{\phi}\right )^{2}
\end{equation}
\end{cor}
We observe that the bound concerns a family of estimators with the same nondiagonal elements.
\subsection{Maximizing the bound}
We want to maximize the right hand side of the inequality. This means to find conditions on a class of measurements that maximize the quantity
\begin{equation}
 C= \left( a(\phi)+b(\phi) -2h_{3r}(r)sin\theta  \int_{\phi+\epsilon - \pi}^{\phi + \epsilon + \pi}\left(\hat{\phi}-\phi\right)(sin\phi x^{12}+cos\phi y^{12})d\hat{\phi}\right )^{2}
\end{equation}
Our purpose is to prove the existence of measurements for which the bound is higher than that of the classical Fisher metric. We impose conditions which the measurements must satisfy and which make them to have a very restricted form. This is only to prove our point, that is to prove that the class of measurements for which there exists a higher bound is not empty. A better approach would be to determine the most general form which has the property of being bounded with this higher bound. The non-negativity of the POVM gives the non-negativity of its eigenvalues, namely
\begin{eqnarray}
\lambda_{1}&=&\frac{1}{2}\left(x^{11}+x^{22}-\sqrt{(x^{11}-x^{22})^{2}+4(x^{12})^{2}+4(y^{12})^{2}}\right) \geq 0\\
\lambda_{2}&=&\frac{1}{2}\left(x^{11}+x^{22}+\sqrt{(x^{11}-x^{22})^{2}+4(x^{12})^{2}+4(y^{12})^{2}}\right) \geq 0
\end{eqnarray}  
The unbiasedness condition of Eq 33 with its LHS changed to $\xi^{i}+\epsilon^{i}$ is
\begin{equation}
 \int_{\phi+\epsilon - \pi}^{\phi + \epsilon + \pi}\left(\hat{\phi}-(\phi+\epsilon)\right)\left(\frac{1+rcos\theta}{2}x^{11}+\frac{1-rcos\theta}{2}x^{22}+rsin\theta\left(cos\phi x^{12}-sin\phi y^{12}\right)\right)d\hat{\phi}\\=0
\end{equation}
where we have made the following general assumptions and specific choices for the matrix elements of the POVM
\begin{equation}
x^{11}(\hat{\phi},\phi+\epsilon)=x^{22}(\hat{\phi},\phi+\epsilon)\geq 0
\end{equation}
\begin{equation}
\int_{\phi+\epsilon - \pi}^{\phi + \epsilon + \pi}x^{11}(\hat{\phi},\phi+\epsilon)d\hat{\phi}= 1 
\end{equation}
\begin{equation}
\int_{\phi+\epsilon - \pi}^{\phi + \epsilon + \pi}x^{12}(\hat{\phi},\phi+\epsilon)d\hat{\phi}=  \int_{\phi+\epsilon - \pi}^{\phi + \epsilon + \pi}y^{12}(\hat{\phi},\phi+\epsilon)d\hat{\phi}= 0 
\end{equation}
We assume further that $x^{11}(\hat{\phi},\phi+\epsilon)$, as a function of $\hat{\phi}$ is symmetric w.r.t. $\phi + \epsilon$. Then if we choose
\begin{equation}
x^{12}=(tan\phi)y^{12}
\end{equation}
the  condition given by Eq 51 is satisfied and we also have
\begin{equation}
sin\phi x^{12}(\hat{\phi},\phi+\epsilon)+cos\phi y^{12}(\hat{\phi},\phi+\epsilon) = \frac{1}{cos\phi}y^{12}(\hat{\phi},\phi+\epsilon)
\end{equation}
Then the quantity to be bounded becomes
\begin{equation}
 C= \left( a(\phi)+b(\phi) -2h_{3r}(r)sin\theta  \int_{\phi+\epsilon - \pi}^{\phi + \epsilon + \pi}\left(\hat{\phi}-\phi\right)\left(      \frac{1}{cos\phi}y^{12}(\hat{\phi},\phi+\epsilon)   \right)d\hat{\phi}\right )^{2}
\end{equation}
With the given assumptions and special conditions a bound for C comes from the nonnegativity of the eigenvalue $\lambda_{1}$. We get
\begin{equation}
x^{11}(\hat{\phi},\phi+\epsilon) \geq \left|\frac{y^{12}(\hat{\phi},\phi+\epsilon)}{cos\phi}\right|
\end{equation}
Assuming further that
\begin{eqnarray}
\frac{y^{12}(\hat{\phi},\phi+\epsilon)}{cos\phi}   \, > \, 0 \quad &for&  \quad \hat{\phi} \, < \, \phi + \epsilon \\
\frac{y^{12}(\hat{\phi},\phi+\epsilon)}{cos\phi}    \, = \, 0 \quad &for&  \quad \hat{\phi} \, = \, \phi + \epsilon \\
\frac{y^{12}(\hat{\phi},\phi+\epsilon)}{cos\phi}   \, < \, 0 \quad &for&  \quad \hat{\phi} \, > \, \phi + \epsilon 
\end{eqnarray}
\begin{equation}
c(\phi)=a(\phi)+b(\phi) \geq 0
\end{equation}

we have
\begin{equation}
 C \leq \left[c(\phi) - 2h_{3r}(r)sin\theta  \left( I_{1}-I_{2}\right)\right]^{2}
\end{equation}
where
\begin{equation}
I_{1}=\int_{\phi+\epsilon - \pi}^{\phi + \epsilon }\left(\left(\hat{\phi}-(\phi+\epsilon)\right)   x^{11}(\hat{\phi},\phi+\epsilon)+\epsilon \frac{1}{cos\phi}y^{12}(\hat{\phi},\phi+\epsilon)\right)d\hat{\phi}
\end{equation}
\begin{equation}
I_{2}=\int_{\phi+\epsilon}^{\phi + \epsilon +\pi}\left(\hat{\phi}-\phi\right)x^{11}(\hat{\phi},\phi+\epsilon)d\hat{\phi}
\end{equation}
Thus for the selected class of measurements we have a lower bound for the variance of the estimator which is different from that given by the SLD. The usefulness of this result is that this bound, for this class of POVMs is higher than not only the bound provided by SLD, which is higher than all those provided by the monotone metrics, but it is higher even from the bound of the corresponding Classical Fisher metric, as it is shown in the next section.
\subsection{An application }
%\subsubsection{Example 1 }
To make the bound explicit we choose the form of $x^{11}$ so that an analytic integration is possible. We take
\begin{equation}
x^{11}(\hat{\phi},\phi+\epsilon)=\frac{1}{3\sqrt{2\pi}erf(\pi/3\sqrt{2})}Exp\left(-\left(\hat{\phi}-(\phi+\epsilon)\right)^{2}/18\right)
\end{equation}
and $y^{12}$ from Eq 58 considered as equality, and with the appropriate  signs.
With this choice we get $a(\phi)=0$ and
\begin{equation}
 b(\phi)=1-\frac{e^{-\pi^{2}/18}\left(-\sqrt{2\pi}+3e^{\pi^{2}/18}erf\left(\frac{\pi}{3\sqrt{2}}\right)-9\left(e^{\pi^{2}/18}-1\right)\sqrt{\frac{2}{\pi}}rsin\theta\right)}{3erf\left(\frac{\pi}{3\sqrt{2}}\right)}
\end{equation}
For the classical Fisher metric for the measurement we have
\begin{equation}
g_{0,Fisher}(r,\theta,\phi)=\int_{\phi-\pi}^{\phi+\pi}\left(\frac{\partial}{\partial \phi}log\left(p_{0}(\hat{\phi}|\phi;r,\theta)\right)    \right)^{2}p_{0}(\hat{\phi}|\phi;r,\theta)d\hat{\phi}
\end{equation}

where
\begin{equation}
p_{0}(\hat{\phi}|\phi;r,\theta)=tr\left[\rho(r,\theta,\phi)\Pi(\hat{\phi},\phi_{0})\right]
\end{equation}
\begin{figure}[htp]
	\centering
\scalebox{0.7}{\includegraphics[angle=0]{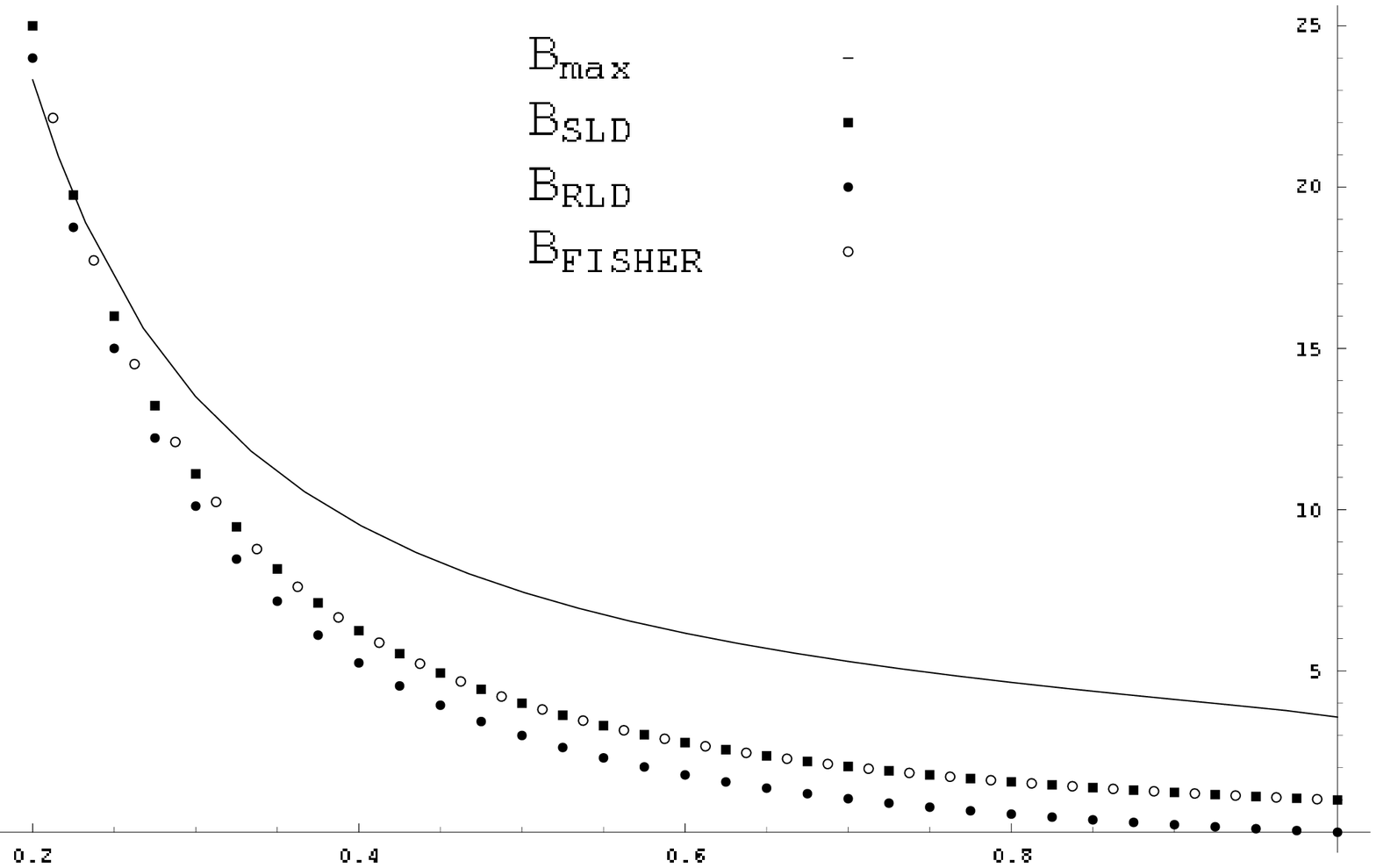}}
%\scalebox{0.3}{\includegraphics{ghi_oik_fig1.pdf}}
%\includegraphics{ghi_oik_fig1.eps}
\caption{Comparison of the different bounds for the variance of the estimator for $\phi$. For the plot, $\phi$ was chosen so that $cos\phi < 0$ and $\theta = \pi /2$, $B_{max}= C_{max}/g$. $B_{SLD}, B_{RLD}, B_{Fisher}$  are the corresponding quantities for SLD, RLD and classical Fisher metric. The x-axis is the parameter r of the state which is related to the puriness of the qubit.}
	\label{fig:comparison}
\end{figure}

Now in order to plot the different bounds we have chosen a value for $\phi$ so that $cos\phi< 0$ and $\theta=\pi/2$.
In Figure 1 we plot $B_{max}= C_{max}/g$ and the corresponding quantities $B_{ SLD}, B_{RLD}, B_{Fisher}$ for SLD, RLD and the classical Fisher metric  versus the parameter r, which is related to the puriness of the state.  We see that the new bound is higher than all the others. The deviation becomes bigger for higher values of r.

\indent \\
\section{Discussion}
We have posed the question whether, for certain classes of quantum estimators the lower bound for the variance of the estimated parameter is higher than those given by the quantum Cramer-Rao inequalities. The current literature deals with questions of saturation of these inequalities, that is which measurements are optimal with respect to their accuracy of estimation. Here we asked whether there exist "bad" measurements for which the error is bounded from bellow with higher bounds. This would be useful in the classification of measurements with respect to their effectiveness. The highest possible bound given by the monotone quantum metrics is that of the Symmetric Logarithmic Derivative. This is majorized by the classical Fisher bound associated with the probability distribution of the measurement. To go beyond these bounds means to construct a metric outside the family of manifestly monotone metrics. To do that we constructed the operator logarithmic derivatives indirectly through the phase space correspondence. That is we mapped the state to the Husimi function, we took the logarithmic derivative and we mapped them back to Hilbert space operators. The related metric is manifestly not monotone. Then using an operator which quantifies the difference of this derivative from the symmetric one we were able to derive a bound for the variance. This depends both on the state and on the measurement. We were not able to determine the general class of the measurements for which the new bound is higher than the previous ones, but using an explicit example we showed that this class is not empty. Now observing that, under the conditions we have imposed on the measurement, we have that one of its eigenvalues is zero, namely $\lambda_{1}=0$, we see that the measurement belongs to the class of sharp measurements \cite{wiseman1}. Further work is needed to determine the class of all measurements which have this property of bounded accuracy and,  most importantly, to analyze  the physical realizability and nature of such measurements. 

%\thanks{One of the authors (F.O.) wishes to thank the Greek State Scholarships Foundation for a research scholarship.}
\appendix
\section{Proof of Theorem 1}
\indent 
The proof is based on similar techniques used in \cite{helstrom1}. First the unbiasedness condition is
\begin{equation}
tr\left(\int \hat{\xi^{j}}\rho(\xi)\Pi(\hat{\xi},\xi+\epsilon)d\hat{\xi}\right)=\xi^{j}
\end{equation} 
Taking the derivative w.r.t. $\xi^{k}$ we have
\begin{equation}
tr\left(\int \hat{\xi^{j}}\rho(\xi)\frac{\partial \Pi(\hat{\xi},\xi+\epsilon)}{\partial\xi^{k}}d\hat{\xi}\right)+tr\left(\int \hat{\xi^{j}}\frac {\partial \rho(\xi)}{\partial\xi^{k}}\Pi(\hat{\xi},\xi+\epsilon)d\hat{\xi}\right)=\delta_{k}^{j}
\end{equation}
On the other hand we have
\begin{equation}
\xi^{j}tr\left(\int \rho(\xi)\frac{\partial \Pi(\hat{\xi},\xi+\epsilon)}{\partial\xi^{k}}d\hat{\xi}\right)+\xi^{j}tr\left(\int \frac {\partial \rho(\xi)}{\partial\xi^{k}}\Pi(\hat{\xi},\xi+\epsilon)d\hat{\xi}\right)=0
\end{equation}
Subtracting these equations we get
\begin{equation}
tr\left(\int \left(\hat{\xi^{j}}-\xi^{j}\right)\frac{\partial \rho(\xi)}{\partial\xi^{k}}\Pi(\hat{\xi},\xi+\epsilon)d\hat{\xi}\right)= B_{k}^{j}+A_{k}^{j}
\end{equation}
or using Equ. 29
\begin{equation}
 tr\left(\int \left(\hat{\xi^{j}}-\xi^{j}\right)\left( h_{k}+\frac{1}{2}(\rho L_{k}+L_{k}\rho)    \right)\Pi(\hat{\xi},\xi+\epsilon)d\hat{\xi}\right)= B_{k}^{j}+A_{k}^{j}
\end{equation}
Now using a Lemma in \cite{helstrom1} we can write this equation as
\begin{eqnarray}
 tr\left(\int \left(\hat{\xi^{j}}-\xi^{j}\right) h_{k}\Pi(\hat{\xi},\xi+\epsilon)d\hat{\xi}\right)+\\
Re\left[tr\left(\int \left(\hat{\xi^{j}}-\xi^{j}\right)\rho L_{k}\Pi(\hat{\xi},\xi+\epsilon)d\hat{\xi}\right)\right]= B_{k}^{j}+A_{k}^{j}
\end{eqnarray}
Multiplying the last equation with the real numbers $y_{j}z_{k}$ and summing over j,k we get
\begin{equation}
 Re \left[tr \int T_{\xi}\rho T_{L} \Pi(\hat{\xi},\xi+\epsilon)d\hat{\xi}\right]=Y^{t}(B+A)Z - tr \int T_{\xi}T_{h} \Pi(\hat{\xi},\xi+\epsilon)d\hat{\xi}
\end{equation}
where
\begin{equation}
T_{L}=\sum_{k=1}^{3}z_{k}L_{k}
\end{equation}
Thus
\begin{equation}
  \left|tr \int \rho T_{L} \Pi(\hat{\xi},\xi+\epsilon)T_{\xi}d\hat{\xi}\right|^{2}\geq \left|Y^{t}(B+A)Z - tr \int T_{\xi}T_{h} \Pi(\hat{\xi},\xi+\epsilon)d\hat{\xi}\right|^{2}
\end{equation}
and from Schwartz inequality we have
\begin{eqnarray}
 \left(tr \int \rho T_{\xi} \Pi(\hat{\xi},\xi+\epsilon)T_{\xi}d\hat{\xi}\right) \left(tr \int \rho T_{L} \Pi(\hat{\xi},\xi+\epsilon)T_{L}d\hat{\xi}\right)  \geq \\\left|Y^{t}(B+A)Z - tr \int T_{\xi}T_{h} \Pi(\hat{\xi},\xi+\epsilon)d\hat{\xi}\right|^{2}
\end{eqnarray}
This gives the inequality that was to be proved.
\\

\end{document}